\documentclass[12pt]{article}
  \usepackage{amsfonts}
  \usepackage{amsmath}
\usepackage{amssymb}
\usepackage{amscd}
\usepackage[dvips]{graphicx}
\begin{document}

~~
\bigskip
\bigskip
\begin{center}
{\Large {\bf{{{Twisted acceleration-enlarged Newton-Hooke space-times and conservative force terms}}}}}
\end{center}
\bigskip
\bigskip
\bigskip
\begin{center}
{{\large ${\rm {Marcin\;Daszkiewicz}}$}}
\end{center}
\bigskip
\begin{center}
\bigskip

{${\rm{Institute\; of\; Theoretical\; Physics}}$}

{${\rm{ University\; of\; Wroclaw\; pl.\; Maxa\; Borna\; 9,\;
50-206\; Wroclaw,\; Poland}}$}

{ ${\rm{ e-mail:\; marcin@ift.uni.wroc.pl}}$}
\end{center}
\bigskip
\bigskip
\bigskip
\bigskip
\bigskip
\bigskip
\bigskip
\bigskip
\bigskip
\begin{abstract}
There are analyzed two classical systems defined on twist-deformed acceleration-enlarged Newton-Hooke
space-times -  nonrelativistic particle moving in constant field force $\vec{F}$ and harmonic oscillator model.
It is demonstrated that only in the case of canonical twist deformation the force terms generated by space-time
noncommutativity remain conservative for both models.
\end{abstract}
\bigskip
\bigskip
\bigskip
\bigskip
\bigskip
\bigskip
{\it ~~~~~~~~~~~~~~~~~~~~~~~~~This paper is dedicated to my daughter Inga}
\eject

Recently, there appeared a lot of papers dealing with noncommutative
classical and quantum  mechanics (see e.g. \cite{mech}-\cite{qmnext})
as well as with field theoretical models (see e.g. \cite{field},
\cite{fieldnext}), in
which  the quantum space-time is employed. The suggestion to use
noncommutative coordinates goes back to Heisenberg and was firstly
formalized by Snyder in \cite{snyder}. Recently, there were also
found formal arguments based mainly on Quantum Gravity \cite{grav1} 
and String Theory models \cite{string1}, 
indicating that space-time at Planck scale  should be
noncommutative, i.e. it should  have a quantum nature. On the other
side, the main reason for such considerations follows from the
suggestion  that relativistic space-time symmetries should be
modified (deformed) at Planck scale, while  the classical Poincare
invariance still remains valid at
larger distances \cite{1a}, \cite{1anext}.

Presently, it is well known, that in accordance with the
Hopf-algebraic classification of all deformations of relativistic
and nonrelativistic symmetries, one can distinguish three basic
types of quantum spaces \cite{class1}, \cite{class2}:\\
\\
{ \bf 1)} Canonical ($\theta^{\mu\nu}$-deformed) space-time
\begin{equation}
[\;{ x}_{\mu},{ x}_{\nu}\;] = i\theta_{\mu\nu}\;\;\;;\;\;\;
\theta_{\mu\nu} = {\rm const}\;, \label{noncomm}
\end{equation}
introduced in  \cite{oeckl}, \cite{chi}
in the case of Poincare quantum group and in \cite{dasz1} 
for its Galilean counterpart.\\
\\
{ \bf 2)} Lie-algebraic modification of classical space
\begin{equation}
[\;{ x}_{\mu},{ x}_{\nu}\;] = i\theta_{\mu\nu}^{\rho}{ x}_{\rho}\;,
\label{noncomm1}
\end{equation}
with  particularly chosen coefficients $\theta_{\mu\nu}^{\rho}$
being constants. This type of noncommutativity has been obtained as
the representations of the $\kappa$-Poincare \cite{kappaP} and
$\kappa$-Galilei
\cite{kappaG} as well as the  twisted relativistic \cite{lie1} 
and  nonrelativistic \cite{dasz1} 
symmetries, respectively. \\
\\
{ \bf 3)} Quadratic deformation of Minkowski and Galilei  space
\begin{equation}
[\;{ x}_{\mu},{ x}_{\nu}\;] = i\theta_{\mu\nu}^{\rho\tau}{
x}_{\rho}{ x}_{\tau}\;, \label{noncomm2}
\end{equation}
with coefficients $\theta_{\mu\nu}^{\rho\tau}$ being constants. This
kind of deformation has been proposed in \cite{qdef}, \cite{paolo},
\cite{lie1}
 at relativistic and in \cite{dasz1} at  nonrelativistic level.\\
\\
Besides, it has been demonstrated in \cite{nh}, that in the case of
so-called acceleration-enlarged Newton-Hooke Hopf algebras
$\,{\mathcal U}_0(\widehat{ NH}_{\pm})$ the (twist) deformation
provides the new  space-time noncommutativity, which is expanding
($\,{\mathcal U}_0(\widehat{ NH}_{+})$) or periodic ($\,{\mathcal
U}_0(\widehat{ NH}_{-})$) in time, i.e. it takes the
form\footnote{The $\,{\mathcal U}_0(\widehat{ NH}_{\pm})$
acceleration-enlarged Newton-Hooke Hopf structures are obtained by
adding to the ${\widehat{NH}}_{\pm}$ algebras (see \cite{lucky0}, \cite{lucky}) the
trivial coproduct $\Delta_{0}(a) = a\otimes 1+ 1\otimes
a$.},\footnote{$x_0 = ct$.}
\begin{equation}
{ \bf 4)}\;\;\;\;\;\;\;\;\;[\;t,{ x}_{i}\;] = 0\;\;\;,\;\;\; [\;{ x}_{i},{ x}_{j}\;] = 
if_{\pm}\left(\frac{t}{\tau}\right)\theta_{ij}(x)
\;, \label{nhspace}
\end{equation}
with time-dependent  functions
$$f_+\left(\frac{t}{\tau}\right) =
f\left(\sinh\left(\frac{t}{\tau}\right),\cosh\left(\frac{t}{\tau}\right)\right)\;\;\;,\;\;\;
f_-\left(\frac{t}{\tau}\right) =
f\left(\sin\left(\frac{t}{\tau}\right),\cos\left(\frac{t}{\tau}\right)\right)\;,$$
and $\theta_{ij}(x) \sim \theta_{ij} = {\rm const}$ or
$\theta_{ij}(x) \sim \theta_{ij}^{k}x_k$. Such a kind  of
noncommutativity  follows from the presence in acceleration-enlarged
Newton-Hooke symmetries $\,{\mathcal U}_0(\widehat{ NH}_{\pm})$ of
the time scale parameter (cosmological constant) $\tau$. As it was
demonstrated in \cite{nh} that  just this parameter   is responsible
for oscillation or expansion of space-time noncommutativity.

It should be noted that both Hopf structures $\,{\mathcal
U}_0(\widehat{ NH}_{\pm})$ contain, apart from rotation $(M_{ij})$,
boost $(K_{i})$ and space-time translation $(P_{i}, H)$ generators,
the additional ones denoted by $F_{i}$, responsible for constant
acceleration. Consequently, if all generators
 $F_{i}$ are equal zero we obtain the twisted Newton-Hooke quantum
 space-times \cite{nh1}, while for time parameter $\tau$ running to infinity
we get the acceleration-enlarged twisted Galilei Hopf structures proposed in
\cite{nh}. In particular,  due to the
presence of generators $F_i$ for $\tau \to \infty$,  we get the new cubic and quartic type of
space-time noncommutativity
\begin{equation}
[\;{ x}_{\mu},{ x}_{\nu}\;] = i\alpha_{\mu\nu}^{\rho_1...\rho_n}{
x}_{\rho_1}...{ x}_{\rho_n}\;, \label{noncomm6}
\end{equation}
with $n=3$ and $4$ respectively, whereas for $F_i\to 0$ and $\tau \to \infty$
 we reproduce the canonical (\ref{noncomm}),
Lie-algebraic (\ref{noncomm1}) and quadratic (\ref{noncomm2})
(twisted) Galilei spaces provided in \cite{dasz1}. Finally, it should be  noted, that all  mentioned above
noncommutative space-times have been  defined as the quantum
representation spaces, so-called Hopf modules (see
\cite{bloch}, \cite{wess}, \cite{oeckl}, \cite{chi}), for quantum acceleration-enlarged
Newton-Hooke Hopf algebras, respectively.

Recently, in the series of papers \cite{romero}-\cite{giri} there has been discussed the impact of different kinds of
space-time noncommutativity on the structure of physical systems. More preciously, it has been demonstrated
that in the case of classical Newtonian models there are generated by quantum spaces additional force terms,
which appear in Newton equation.  Such an observation
permitted to analyze the Pioneer anomaly phenomena \cite{toporzelek} with use of the classical nonrelativistic particle model
defined on $\kappa$-Galilei quantum space-time \cite{kappaG}. Besides, there has been suggested in \cite{noninertial}, that deformation of
classical systems can be identified with their noninertial transformation, while the forces of
inertia should be identical with force terms produced by space-time noncommutativity.

In this article, we check which forces generated by mentioned above acceleration-enlarged Newton-Hooke quantum
spaces remain conservative. We perform our investigations in context of two simplest physical systems -
nonrelativistic particle moving in constant external field force $\vec{F}$ and  classical oscillator model.
Besides, it should be  noted that we consider only noncommutative space-times equipped with classical time and quantum spatial
directions, i.e. we consider  spaces of the form
\begin{equation}
[\;t,{ x}_{i}\;] = 0\;\;\;,\;\;\; [\;{ x}_{1},{ x}_{2}\;] =
if({t})\;\;\;,\;\;\;[\;x_1,{ x}_{3}\;] = 0 =[\;x_2,{ x}_{3}\;]\;\;;\;\;i=1,2,3
\;, \label{spaces}
\end{equation}
with function $f({t})$ given by
\begin{eqnarray}
f({t})&=&f_{\kappa_1}({t}) =
f_{\pm,\kappa_1}\left(\frac{t}{\tau}\right) = \kappa_1\,C_{\pm}^2
\left(\frac{t}{\tau}\right)\;, \label{w2}\\
f({t})&=&f_{\kappa_2}({t}) =
f_{\pm,\kappa_2}\left(\frac{t}{\tau}\right) =\kappa_2\tau\, C_{\pm}
\left(\frac{t}{\tau}\right)S_{\pm} \left(\frac{t}{\tau}\right) \;,
\label{w3}\\
f({t})&=&f_{\kappa_3}({t}) =
f_{\pm,\kappa_3}\left(\frac{t}{\tau}\right) =\kappa_3\tau^2\,
S_{\pm}^2 \left(\frac{t}{\tau}\right) \;, \label{w4}\\
f({t})&=&f_{\kappa_4}({t}) =
 f_{\pm,\kappa_4}\left(\frac{t}{\tau}\right) = 4\kappa_4
 \tau^4\left(C_{\pm}\left(\frac{t}{\tau}\right)
-1\right)^2 \;, \label{w5}\\
f({t})&=&f_{\kappa_5}({t}) =
f_{\pm,\kappa_5}\left(\frac{t}{\tau}\right) = \pm \kappa_5\tau^2
\left(C_{\pm}\left(\frac{t}{\tau}\right)
-1\right)C_{\pm} \left(\frac{t}{\tau}\right)\;, \label{w6}\\
f({t})&=&f_{\kappa_6}({t}) =
f_{\pm,\kappa_6}\left(\frac{t}{\tau}\right) = \pm \kappa_6\tau^3
\left(C_{\pm}\left(\frac{t}{\tau}\right) -1\right)S_{\pm}
\left(\frac{t}{\tau}\right)\;; \label{w7}
\end{eqnarray}
$$C_{+/-} \left(\frac{t}{\tau}\right) = \cosh/\cos \left(\frac{t}{\tau}\right)\;\;\;{\rm and}\;\;\;
S_{+/-} \left(\frac{t}{\tau}\right) = \sinh/\sin
\left(\frac{t}{\tau}\right) \;.$$
As it was already mentioned, in $\tau \to \infty$ limit the above quantum spaces reproduce the canonical (\ref{noncomm}),
Lie-algebraic (\ref{noncomm1}), quadratic (\ref{noncomm2}) as  well as cubic and quartic (\ref{noncomm6}) type of
space-time noncommutativity, with\footnote{Space-times (\ref{nw2})-(\ref{nw4}) correspond to the
twisted Galilei Hopf algebras provided in \cite{dasz1}, while the quantum spaces (\ref{nw5})-(\ref{nw7}) are
associated with acceleration-enlarged Galilei Hopf structures \cite{nh}.} 
\begin{eqnarray}
f_{\kappa_1}({t}) &=& \kappa_1\;,\label{nw2}\\
f_{\kappa_2}({t}) &=& \kappa_2\,t\;,\label{nw3}\\
f_{\kappa_3}({t}) &=& \kappa_2\,t^2\;,\label{nw4}\\
f_{\kappa_4}({t}) &=& \kappa_4\,t^4\;,  \label{nw5}\\
f_{\kappa_5}({t}) &=& \frac{1}{2}\kappa_5\,t^2\;, \label{nw6}\\
f_{\kappa_6}({t}) &=& \frac{1}{2}\kappa_6\,t^3\;. \label{nw7}
\end{eqnarray}
Of course, for all  parameters $\kappa_a$ running to zero the above deformations disappear.

Let us now turn to the mentioned above dynamical models. Firstly, we start with
following phase space\footnote{We use the correspondence relation $\{\;a,b\;\}
= \frac{1}{i}[\;\hat{a},\hat{b}\;]$  $(\hbar = 1)$.}
\begin{equation}
\{\;t,{ {\bar x}}_{i}\;\} = 0 \;\;\;,\;\;\;\{\;{ {\bar x}}_{1},{ {\bar
x}}_{2}\;\}  = f_{\kappa_a}(t)\;\;\;,\;\;\;
\{\;{ {\bar x}}_{1},{{\bar
x}}_{3}\;\}= 0=\{\;{ {\bar x}}_{2},{ {\bar
x}}_{3}\;\} \;, \label{beyond}
\end{equation}
\begin{equation}
\{\;{ {\bar x}}_{i},{\bar p}_j\;\} = \delta_{ij}\;\;\;,\;\;\;\{\;{
{\bar p}}_{i},{ {\bar p}}_{j}\;\} = 0\;, \label{genin2a}
\end{equation}
corresponding to the quantum space-time (\ref{spaces}). One can check that the  relations (\ref{beyond}),
(\ref{genin2a}) satisfy the Jacobi identity and for deformation
parameters $\kappa_a$ running to zero become classical. Next, we define the Hamiltonian function for
nonrelativistic particle moving in constant field force $\vec{F}$ as follows
\begin{eqnarray}
H(\bar{p},\bar{x}) = \frac{1}{2m}\left({\bar p}_{1}^2 +
{\bar p}_{2}^2 + {\bar p}_{3}^2\right) -  \sum_{i=1}^{3}F_i \bar{x}_i\;.\label{ham1}
\end{eqnarray}
In order to analyze the above system we represent the
noncommutative variables $({\bar x}_i, {\bar p}_i)$ on classical
phase space $({ x}_i, { p}_i)$ as  (see e.g. \cite{lumom},
\cite{kijanka}, \cite{giri})
\begin{equation}
{\bar x}_{1} = { x}_{1} - \frac{f_{\kappa_a}(t)}{2}
p_2\;\;\;,\;\;\;{\bar x}_{2} = { x}_{2} +\frac{f_{\kappa_a}(t)}{2}
p_1\;\;\;,\;\;\; {\bar x}_{3}= x_3 \;\;\;,\;\;\; {\bar p}_{i}=
p_i\;, \label{rep}
\end{equation}
where
\begin{equation}
\{\;x_i,x_j\;\} = 0 =\{\;p_i,p_j\;\}\;\;\;,\;\;\; \{\;x_i,p_j\;\}
=\delta_{ij}\;. \label{classpoisson}
\end{equation}
Then, the  Hamiltonian (\ref{ham1}) takes the form
\begin{eqnarray}
{{H}}({ p},{ x})=H_f(t) = \frac{1}{2m}\left({ p}_{1}^2 +
{ p}_{2}^2 + { p}_{3}^2\right) - \sum_{i=1}^{3}F_i x_i + F_1 \frac{f_{\kappa_a}(t)}{2}p_2 -
F_2 \frac{f_{\kappa_a}(t)}{2}p_1 \label{hamoscnew}\;.
\end{eqnarray}
Using the formulas (\ref{classpoisson}) and (\ref{hamoscnew}) one gets
the following canonical Hamiltonian equations of motions $(\dot{{o}}_i = \frac{d}{dt}o_i = \{\;o_i,H\;\})$
\begin{eqnarray}
&&\dot{x}_{1} = \frac{p_1}{m} - \frac{f_{\kappa_a}(t)
}{2}F_2 \;\;\;,\;\;\; \dot{p}_{1} = F_1
\;,\label{ham1a}\\
 &~~&~\cr
&&\dot{x}_{2} = \frac{p_2}{m} + \frac{f_{\kappa_a}(t)
}{2}F_1\;\;\;,\;\;\; \dot{p}_{2} = F_2
\;,\label{ham2a}\\
&~~&~\cr
&&~~~~~\dot{x}_{3} = \frac{p_3}{m}\;\;\;,\;\;\;\dot{p}_{3} =
F_3\;,\label{ham3a}
\end{eqnarray}
which when combined yield the equations
\begin{equation}
\left\{\begin{array}{rcl} m\ddot{x}_1  &=&{F_1} - \frac{m\dot{f}_{\kappa_a}(t)
}{2}F_2 = G_1(t)\\
 &~~&~\cr
 m\ddot{x}_2  &=&
{F_2} + \frac{m\dot{f}_{\kappa_a}(t)
}{2}F_1 = G_2(t)\\
 &~~&~\cr
 m\ddot{x}_3  &=& {F_3} = G_2
 \;.\end{array}\right.\label{dddmixednewton1}
\end{equation}
First of all, by trivial integration one can find the solution of  above system; it looks as follows
 \begin{eqnarray}
&&{x}_{1}(t) = \frac{F_1}{2m}t^2  + v^0_1t - \frac{F_2}{2}\int_{0}^{t}{f}_{\kappa_a}(t')dt'
\;\;\;,\;\;\;{x}_{2}(t) = \frac{F_2}{2m}t^2  + v^0_2t + \frac{F_1}{2}\int_{0}^{t}{f}_{\kappa_a}(t')dt'
\nonumber\\
&&~\cr
&&~~~~~~~~~~~~~~~~~~~~~~~~~~~~~~~{x}_{3}(t) = \frac{1}{2m}F_3t^2  + v^0_2t +x^0_3\;,\label{sol3}
\end{eqnarray}
with $v^0_i$ and $x_3^0$ denoting the initial velocity and position of particle respectively.
Further, one should  observe that the noncommutativity (\ref{spaces}) generates the new, time-dependent force term
$\vec{G}(t) = \left[\;G_1(t),G_2(t),G_3\;\right]$, which for deformation parameters $\kappa_a$ approaching
zero reproduces undeformed force $\vec{F}$. Besides, it should be noted that for $f(t) = \kappa_1 = \theta$ and $f(t) = \kappa_2 t$ (see formulas (\ref{nw2}) and (\ref{nw3}) respectively)
we recover two models provided in \cite{daszwal}. First of them does not introduce any modification of Newton equation, while
the second one generates the constant acceleration of particle. Finally, let us notice that for arbitrary function
$f_{\kappa_a}(t)$ the rotation of force $\vec{G}(t)$ vanishes
\begin{eqnarray}
\vec{\nabla}\times\vec{G}(t) = 0\;,
\end{eqnarray}
i.e. the generated by space-time noncommutativity (\ref{spaces}) force term (\ref{dddmixednewton1})  remains conservative, and
 the corresponding (nonstationary) potential function takes the form\footnote{$\vec{G}(t) = - {\rm grad}\, V(\vec{x},t)$.}
\begin{eqnarray}
V(\vec{x},t) = -\sum_{i=1}^{3}F_i x_i - \frac{\dot{f}_{\kappa_a}(t)}{2} \left(F_1x_2 - F_2x_1\right)\;.\label{newpot1}
\end{eqnarray}

Let us now turn to  the second dynamical system - to the harmonic oscillator model described  by
\begin{equation}
{{H}}({\bar p},{\bar x})=\frac{1}{2m}\left({\bar p}_{1}^2 +
{\bar p}_{2}^2 + {\bar p}_{3}^2\right) +
\frac{m\omega^2}{2}\left({\bar x}_{1}^2 + {\bar x}_{2}^2 +
{\bar x}_{3}^2\right)\;. \label{hamosc}
\end{equation}
where $m$  and  $\omega$ denote the mass of particle and frequency of oscillation respectively. Using transformation
rules (\ref{rep}) one can rewrite the above Hamiltonian function in terms of commutative variables as follows
\begin{eqnarray}
H_f(t) =\frac{\left({ p}_{1}^2 + { p}_{2}^2\right)}{2M_f(t)}
+\frac{m\omega^2}{2}\left({ x}_{1}^2 +
{x}_{2}^2\right)+\frac{f_{\kappa_a}(t)m\omega^2L_3}{2} +\frac{{
p}_{3}^2}{2m} +\frac{m\omega^2{ x}_{3}^2}{2}\;, \label{genhamoscnew}
\end{eqnarray}
with
\begin{equation}
M_f(t)= \frac{m}{1 + \frac{m^2\omega^2}{4}f_{\kappa_a}^2(t) }\;\;\;,\;\;\;L_3 = x_1 p_2 - x_2 p_1 \;.
\label{genmassfre}
\end{equation}
Next, we find the corresponding Newton equation  
which takes the form \cite{oscylator}
\begin{equation}
\left\{\begin{array}{rcl} m\ddot{x}_1 &=& \dfrac{m^2\omega^2
f_{\kappa_a}(t)}{2}\left( \dot{f}_{\kappa_a}(t)M_f(t) \dot{x}_1
+2\dot{x}_2\right)+\\
&~~&~\cr &+& \dfrac{m^2\omega^2\dot{f}_{\kappa_a}(t)}{2}
\left(1-\dfrac{m\omega^2 M_f(t)}{2}f^2_{\kappa_a}(t)\right)x_2-m\omega^2 x_1 = H_1(\vec{x},{\dot{\vec{x}}},t)\\
 &~~&~\cr
 m\ddot{x}_2  &=&
\dfrac{m^2\omega^2 f_{\kappa_a}(t)}{2}\left( \dot{f}_{\kappa_a}(t)M_f(t)
\dot{x}_2 -2\dot{x}_1\right)+\\
&~~&~\cr &+& \dfrac{m^2\omega^2\dot{f}_{\kappa_a}(t)}{2}
\left(\dfrac{m\omega^2 M_f(t)}{2}f^2_{\kappa_a}(t)-1\right)x_1-m\omega^2 x_2 = H_2(\vec{x},{\dot{\vec{x}}},t)\\
 &~~&~\cr
 m\ddot{x}_3  &=&-m\omega^2 x_3 = H_3(\vec{x})
 \;.\end{array}\right.\label{gendddmixednewton1}
\end{equation}
Firstly, it should be noted that
the solution of the above system has been studied numerically in \cite{oscylator} only for  most simple (canonical) case. Besides, one can  observe that
for function $f(t)$ approaching zero the discussed  deformation disappears.
Finally, by simple calculation  one can find the rotation of  generated by space-time noncommutativity (\ref{spaces}) force $\vec{H}(\vec{x},{\dot{\vec{x}}},t)$; it looks as follows
\begin{eqnarray}
\vec{\nabla}\times\vec{H}(\vec{x},{\dot{\vec{x}}},t) = {\rm {\hat e}_3}m^2\omega^2 \dot{f}_{\kappa_a}(t)\left[\frac{m\omega^2M_f(t)}{2}f_{\kappa_a}^2(t)-1\right]\;.
\end{eqnarray}
It is easy to see, that  r.h.s. of the above identity vanishes only for canonical deformation (\ref{nw2}),  and 
then, the generated force term takes the form
\begin{eqnarray}
H_1(\vec{x},{\dot{\vec{x}}}) &=& -m\omega^2 x_1 + m^2\omega^2\theta \dot{x}_2\;,\label{f1}\\
H_2(\vec{x},{\dot{\vec{x}}}) &=& -m\omega^2 x_2 - m^2\omega^2\theta \dot{x}_1\;,\label{f2}\\
H_3(\vec{x}) &=& -m\omega^2 x_3 \;.\label{f3}
\end{eqnarray}
Of course, in the case of remaining spaces the obtained forces  become nonconservative.

Let us summarize our results. In this article we investigate the simple property of force terms generated by different types of quantum spaces, i.e. we check
which of them remain conservative. We perform our investigations in context of two basic systems:  noncommutative
particle moving in  external constant field force $\vec{F}$ and  harmonic oscillator model. Particularly, we demonstrate that in the case of first dynamical system all considered quantum space-times produce conservative force terms, while for the second model  such a situation appears only for the canonical type of noncommutativity. This result confirms that the
canonical deformation is close to the undeformed one, i.e.  for canonically deformed quantum  space both analyzed models do not change the conservative nature of its dynamics. Finally, it should be noted that obtained results describe only two basic (mentioned above) classical
systems. Obviously,  our kind of investigations can be applied to much more complicated physical models. However, the main aim
of this article is only to signalize and to illustrate  an interesting problem, and the choice of such simple systems is dictated by technical transparency  of performed  calculations.

 \section*{Acknowledgments}
The author would like to thank J. Lukierski
for valuable remarks.\\
This paper has been financially supported by Polish Ministry of
Science and Higher Education grant NN202318534.


\begin{thebibliography}{99}
\bibitem{mech}A. Deriglazov, JHEP 0303, 021 (2003); hep-th/0211105
\bibitem{mechnext}S. Ghosh, Phys. Lett. B 648, 262
(2007)
\bibitem{qmnext}M. Chaichian, M.M. Sheikh-Jabbari, A. Tureanu, Phys.
Rev. Lett. 86, 2716 (2001)
\bibitem{field}M. Chaichian, P. Pre\v{s}najder and  A. Tureanu,
Phys. Rev. Lett. 94, 151602 (2005)
\bibitem{fieldnext}G. Fiore, J. Wess, Phys. Rev. D
75, 105022 (2007)
\bibitem{snyder}H.S. Snyder, Phys. Rev. 72, 68 (1947)
\bibitem{grav1}S. Doplicher, K. Fredenhagen, J.E. Roberts, Phys. Lett. B 331, 39
(1994)
\bibitem{string1}A. Connes, M.R. Douglas, A. Schwarz, JHEP 9802, 003
(1998)
\bibitem{1a}S. Coleman, S.L. Glashow, Phys. Rev. D 59, 116008
(1999)
\bibitem{1anext}
R.J. Protheore, H. Meyer, Phys. Lett. B 493, 1 (2000)
\bibitem{class1}S. Zakrzewski, \textit{"Poisson Structures on the Poincare
group"}
; q-alg/9602001
\bibitem{class2}
Y. Brihaye, E. Kowalczyk, P. Maslanka, \textit{"Poisson-Lie structure on Galilei
group"}
; math/0006167
\bibitem{oeckl}R. Oeckl, J. Math. Phys. 40, 3588 (1999)
\bibitem{chi}M. Chaichian, P.P. Kulish, K. Nashijima, A. Tureanu, Phys. Lett. B
604, 98 (2004)
\bibitem{dasz1}M. Daszkiewicz,
Mod. Phys. Lett. A 23, 505 (2008); arXiv: 0801.1206
[hep-th]
\bibitem{kappaP}J. Lukierski, A. Nowicki, H. Ruegg and V.N. Tolstoy, Phys. Lett.
B 264, 331 (1991)
\bibitem{kappaG}S. Giller, P. Kosinski, M. Majewski, P. Maslanka
and J. Kunz, Phys. Lett. B 286, 57 (1992)
\bibitem{lie1}
J. Lukierski and M. Woronowicz, Phys. Lett. B 633, 116 (2006); hep-th/0508083
\bibitem{qdef}O. Ogievetsky, W.B.  Schmidke, J. Wess, B. Zumino, Comm. Math. Phys.
150, 495 (1992)
\bibitem{paolo}
P. Aschieri, L. Castellani, A.M. Scarfone, Eur. Phys. J. C 7, 159
(1999)
\bibitem{nh}M. Daszkiewicz, Acta Phys. Pol. B 41, 1889 (2010); arXiv: 1007.4654 [math-ph], [hep-th]
\bibitem{lucky0}J. Lukierski, P.C. Stichel, W.J. Zakrzewski, Eur. Phys. J. C 55, 119 (2008);
arXiv: 0710.3093 [hep-th]
\bibitem{lucky}J. Gomis, J. Lukierski, Phys. Lett. B 664, 107
(2008); arXiv: 0803.3965 [hep-th]
\bibitem{nh1}M. Daszkiewicz, Mod. Phys. Lett. A 24, 1325 (2009); arXiv: 0904.0432 [hep-th]
\bibitem{bloch}C. Blohmann, J. Math. Phys. 44, 4736 (2003); q-alg/0209180
\bibitem{wess}J. Wess, \textit{"Deformed coordinate spaces: Derivatives"}
; hep-th/0408080
\bibitem{romero}J.M. Romero and J.D. Vergara, Mod. Phys. Lett. A 18,
1673 (2003);  hep-th/0303064
\bibitem{romero1}
J.M. Romero, J.A. Santiago, J.D. Vergara, Phys. Lett. A 310, 9
(2003); hep-th/0211165
\bibitem{cytowania}Y. Miao, X. Wang, S. Yu,
\textit{"Classical mechanics on noncommutative space with
Lie-algebraic structure"}; arXiv: 0911.5227 [math-ph]
\bibitem{toporzelek} E. Harikumar, A.K. Kapoor,
\textit{"Newton equation on the kappa  space-time and the Kepler
problem"}; arXiv: 1003.4603 [hep-th]
\bibitem{daszwal}M. Daszkiewicz, C.J. Walczyk, Phys. Rev. D 77, 105008 (2008); 0802.3575 [mat-ph], [hep-th]
\bibitem{oscylator}M. Daszkiewicz, C.J. Walczyk, Acta Phys. Pol. B 40, 293
(2009); arXiv: 0812.1264 [hep-th]
\bibitem{noninertial}M. Daszkiewicz, Acta Phys. Pol. B 41, 1881 (2010); arXiv: 1007.4656 [math-ph], [hep-th]
\bibitem{lumom}J. Lukierski, P.C. Stichel, W.J. Zakrzewski, Annals of Phys.
260, 224 (1997); hep-th/9612017
\bibitem{kijanka}A. Kijanka, P. Kosinski, Phys. Rev. D 70, 12702
(2004); hep-th/0407246
\bibitem{giri}P.R. Giri, P. Roy, \textit{"Noncommutative oscillator, symmetry and Landau
problem"}; 0803.4090 [hep-th]
\end{thebibliography}
\end{document}